\documentclass{egpubl}
\usepackage{eg2024}
 
\ConferenceSubmission   
\usepackage[T1]{fontenc}
\usepackage{dfadobe}  

\usepackage{cite}  
\BibtexOrBiblatex
\electronicVersion
\PrintedOrElectronic
\ifpdf \usepackage[pdftex]{graphicx} \pdfcompresslevel=9
\else \usepackage[dvips]{graphicx} \fi

\usepackage{egweblnk} 
\usepackage{lineno}

\usepackage[algo2e,ruled,linesnumbered]{algorithm2e}
\usepackage{booktabs}   
\usepackage{subcaption} 
\usepackage{amsmath}   
\usepackage{caption}
\usepackage{listings}
\usepackage{xcolor}
\usepackage{graphicx}
\usepackage{xspace}
\usepackage{algorithm}
\usepackage{algpseudocode}
\usepackage{algcompatible}
\usepackage{multirow}
\usepackage{centernot}
\usepackage{amsthm}

\newtheorem{reduction}{Reduction}
\usepackage{enumitem} 

\usepackage[T1]{fontenc}
\usepackage{titlesec}
\titleformat{\paragraph}[runin]
{\bfseries}{\theparagraph}{1em}{}

\algnewcommand{\algorithmicsubalgorithm}[1]{\textbf{#1}}
\algdef{SE}[SUBALG]{SubAlgorithm}{EndSubAlgorithm}{\algorithmicsubalgorithm}{\algorithmicend\ \algorithmicsubalgorithm}
\algtext*{EndSubAlgorithm}
\algtext*{ENDIF}

\algrenewcommand\algorithmiccomment[1]{\hfill \(\triangleright\) #1}

\AtBeginDocument{%
  \providecommand\BibTeX{{%
    \normalfont B\kern-0.5em{\scshape i\kern-0.25em b}\kern-0.8em\TeX}}}

\title{Mochi: Fast \& Exact Collision Detection}

\author[D. Mandarapu]
{\parbox{\textwidth}{\centering Durga Keerthi Mandarapu,  Nicholas James, \& Milind Kulkarni\\ Purdue University}}

\begin{document}


\maketitle
\begin{abstract}
    Collision Detection (CD) has several applications across the domains such as robotics, visual graphics, and fluid mechanics.
    Finding exact collisions between the objects in the scene is quite computationally intensive.
    To quickly filter the object pairs that do not result in a collision, bounding boxes are built on the objects, indexed using a Bounding Volume Hierarchy(BVH), and tested for intersection before performing the expensive object-object intersection tests.
    In state-of-the-art CD libraries, accelerators such as GPUs are used to accelerate BVH traversal by building specialized data structures.
    The recent addition of ray tracing architecture to GPU hardware is designed to do the same but in the context of implementing a Ray Tracing algorithm to render a graphical scene in real-time.
    We present Mochi, a fast and exact collision detection engine that accelerates both the broad and narrow phases by taking advantage of the capabilities of Ray Tracing cores.
    We introduce multiple new reductions to perform generic CD to support three types of objects for CD: simple spherical particles, objects describable by mathematical equations, and complex objects composed of a triangle mesh.
    By implementing our reductions, Mochi achieves several orders of magnitude speedups on synthetic datasets and 5x-28x speedups on real-world triangle mesh datasets.
    We further evaluate our reductions thoroughly and provide several architectural insights on the ray tracing cores that are otherwise unknown due to their proprietorship.


\end{abstract}  
\section{Introduction}
Collision Detection (CD) plays a significant part in several applications such as fluid mechanics\cite{breuer2015modeling}, computer graphics\cite{oibvh}, virtual reality\cite{cordier2002real}, and path planning\cite{yershova2007improving}, for simulating real-world object behaviors\cite{weller2013new}.
The CD kernel takes a scene with multiple objects as input and determines all the pairs of colliding objects along with any application-specific parameters such as point of collision, time of collision, or the volume displaced due to collision. 

Exhaustively searching all pairs of objects for collisions takes $O(n^2)$ time. 
Hence, the CD pipeline employs a \textit{broad phase} before the \textit{narrow phase}.
In the broad phase of detecting collisions, objects or parts of the object that may not cause a collision are quickly removed by employing techniques such as sweep-and-prune (SAP), Bounding Volume Hierarchy (BVH), and spatial hashing.
The potential pairs of colliding objects are then tested using expensive object intersection tests to determine the exact pairs of colliding objects in the narrow phase.
The computational intensity of finding the exact collisions led the community to explore the broad phase more than the narrow phase.
However, knowing if the collision happened or will happen precisely in several scenarios is essential.
For example, are parts of a mechanical system breaking on impact or self-colliding and building up stress to eventually cause fatigue failure?
To address these concerns, we present {\em Mochi, a fast and exact collision detection engine} that accelerates both the broad and narrow phases with the help of Ray Tracing (RT) cores.

Objects in the real world have various shapes, and so they are bounded by a box and indexed. 
BVH is a spatial tree that indexes objects' bounding volume (BV).
BVH efficiently culls non-intersection space because it allows to detect all the collisions in $O(n\cdot logn)$ time that it takes to traverse the BVH.
The faster and more efficiently the BVH can be traversed, the faster the simulations can end, the quicker the graphical rendering can be done, and the sooner the precautions can be taken to avoid collisions (Section~\ref{sec:backgorund_cd} describes this in detail).

In state-of-the-art CD libraries, accelerators such as GPUs are used to implement specialized BVHs and traversal techniques to detect collisions.
The recent addition of ray tracing architecture (RTX) to GPU hardware enables programmers to pass on the BVH operation required for implementing the RT algorithm to render a graphical scene in real-time.
In the RT algorithm, one or more rays are cast per pixel along the camera's direction to obtain information on rendering the particular pixel.
While AMD \cite{amd} and Intel \cite{intel} GPUs have hardware support for ray tracing, Nvidia Turing and later generation RTX GPUs are designed with specialized RT cores \cite{nvidia_white}.
These RT cores build BVH with the objects in the scene and return all the intersections between the rays and the objects.

The RT cores build and traverse BVH, which is just the functionality needed to solve CD.
Nonetheless, it is the traditional implementation of the CD algorithm that poses several challenges for the mapping:
\begin{enumerate}
    \item Once BVH is built, all we have to do to find the collisions is to traverse the BVH and check for the overlapping bounding volumes (BVs).
    However, we can not access the BVH nor control the traversal like the previous GPU-based CD solutions\cite{oibvh, wang2018} since access to the BVH tree is restricted by the RT hardware.
    
    \item RT cores only provide us with ray-BV or ray-triangle intersection tests. However, objects in the real world are complex, involving collisions with themselves or several parts of another complex object. 
    Exact CD involves precise collision/intersection tests between these objects, and it is not straightforward to implement such tests on RT hardware. 
    
    \item RT cores are not general-purpose compute units like shader cores on GPUs.
    To hand off a non-RT application like CD to threads on these application-specific chips, we must present the application in terms of ray-tracing-specific operations.
\end{enumerate}

To investigate a solution for these challenges, we turn to the prior work on generalizing these cores, which includes computing force-directed graph drawings of an $n$-body system~\cite{zellman}, performing and optimizing nearest neighbor search~\cite{evangelou,trueknn,rtnn}, and computing clusters using DBSCAN algorithm~\cite{dbscan}.
All of these works use a particular reduction in different contexts: bound each data point in space with a sphere and shoot an infinitesimal-length ray from the center of each sphere to learn of the points in the neighborhood in the form of ray-sphere intersections.
We will refer to this reduction as RT-NNS throughout the paper, in short for Ray Tracing Nearest Neighbor Search (Details in Section~\ref{sec:uniformradius}). 

RT-NNS reduction can be used to perform CD with a bit of tweaking. 
However, the objects in the scene will be limited to uniform radius spheres.
Moreover, RT-NNS reduction uses an infinitesimal-length ray so that the intersected objects also contain the source of the ray; this means a significant overlap of space between the intersected object and the object where the ray originated.
However, in case of collisions, the colliding objects touch at the boundaries if they are rigid, or they could move into each others' previous positions if they are deformable.
Trying to solve CD along the lines of RT-NNS reduction results in overlapping BVs by construction. This degrades BVH's quality since BVH's performance thrives on how efficiently we can choose non-overlapping BVs for the objects. 
For the reasons above, {\em we introduce multiple new reductions to perform generic CD.}

Mochi leverages a crucial fact about the {\em collides with} relation: that it is commutative. That is, if an object $a$ collides with an object $b$, then $b$ also collides with $a$. Hence, it is not necessary for each object to precisely detect all of the other objects that it collides with. Instead, Mochi need only ensure that {\em one} of a pair of colliding objects detects the collision. Mochi thus
detects collisions of these objects by tracing their periphery in space by shooting several rays of finite length. 
As long as all the objects have convex surfaces, tracing is sufficient to say at least one of the rays scouting the boundaries of the colliding pair reports the collision.
Mochi identifies and supports three types of objects for CD: spherical particle, object describable by a mathematical equation, and complex object composed of a triangle mesh.

\paragraph*{Our contributions}
In this paper, {\em we map collision detection, which is the problem of BVH construction and traversal, to a ray tracing problem handled by RT cores.} 
Our main contributions are as follows.
\begin{enumerate}
    \item We first extend RT-NNS reduction to support spheres of uniform radius by using larger representative spheres.
    Then, present a different reduction that helps Mochi detect collisions between spheres with different radii.
    \item We further show how to perform generic object-object intersection tests on an architecture that only supports ray-object intersections as long as a mathematical equation represents the objects.
    \item To detect collisions between a triangle mesh, including self-collisions, we present our final reduction, which extends the ability of RT cores that perform only non-coplanar ray-triangle intersection tests to both coplanar and non-coplanar triangle-triangle intersection tests.
\end{enumerate}

Section \ref{sec:design} describes all our reductions and workings of Mochi in greater detail. 
We discuss the implementation details and practical challenges in section \ref{sec:implementation}.
We thoroughly evaluate our reductions in section \ref{sec:evaluation} and provide several architectural insights on the RT-cores that are otherwise unknown due to their proprietorship.

\section{Background}
\label{sec:background}
\subsection{Collision Detection}
\label{sec:backgorund_cd}
Collision detection is highly used in visualization graphics and robotics.
To simulate real-world environments in a computer, objects are represented by a geometric model, such as polygonal surfaces.
To find collisions between such complex geometries is a time-consuming and complicated task.
To employ optimizations and acceleration structures, the collision detection pipeline is divided into two phases - broad and narrow.
The purpose of the broad phase is to be quicker than efficient in weeding out the pairs of objects that do not result in a collision.
It is the bounding volumes of the objects that are checked for collision in this phase rather than the objects themselves.
An efficient broad-phase filtering technique would ideally detect an optimal number of colliding pair candidates so that fewer expensive object intersection tests can be done in the narrow phase.
In applications like robotics and autonomous vehicles \cite{chakravarthy1998obstacle}, CD is used to avoid obstacles rather than finding them after the effect.
For such time-sensitive applications, it is not enough to just accelerate the CD pipeline. The preparation process of building the BVH on the objects also needs to be faster.
As we show in evaluation \ref{sec:evaluation}, RT cores help us in building BVH faster than state-of-the-art techniques.

\subsection{Related Work on Collision Detection}
Bullet \cite{bullet} is a popular collision detection library that supports moving bodies' dynamics in games and robots.
Though Bullet is a perfect baseline for particle collision detection, its extensive code base is developed for CPU, and its GPU version is not well optimized.
There is extensive research on accelerating and parallelizing BVH-based CD on GPU.
Lauterbach et al. \cite{lauterbach2010gproximity} is one of the initial works to use GPU cores to perform collision queries on rigid and deformable bodies. 
They implement parallel BVH traversal by constructing a tree front, the list of nodes visited.
The front-based traversal has been greatly optimized by several works, with the recent one being Wang et al. \cite{wang2018}.
They use histogram sort and tree front log to build a cache-friendly layout of tree front and BVH.
They also use the front log to restructure the BVH when its quality drops.
Chitalu et al. \cite{oibvh} further optimize the layouts of the accelerating structures by building a specialized BVH called binary ostensibly implicit trees and speed up the access of BVH nodes during the traversal.

\subsection{Ray Tracing Cores}
Ray tracing is a graphics rendering algorithm that depicts the objects in a 3D scene on 2D visual surfaces.
Several rays are modeled with the camera position as the source and traced till they either hit or miss the objects in the scene.
Based on the properties of the object the ray hits, several optical effects are rendered, for example, reflection, blurring, illumination, etc.
The rays can spawn new rays to improve the visualization.
To accelerate the ray-object hits, the objects are put inside an axis-aligned bounding box (AABB), and a tree called bounding volume hierarchy (BVH) is constructed by hierarchical grouping the bounding boxes.
With the goal of accelerating ray tracing, RT cores build BVH internally and provide an interface to traverse through launching a ray.
RT cores can construct BVH only on triangles or AABBS as the nodes. 
The ray tracing architecture hands off certain operations to shader cores, especially when the nodes are AABBs.

\subsubsection{Programming Model} 
Optix provides an interface to both the CUDA/shader cores and RT cores, allowing the users to make use of 
not only the ray tracing architecture but also traditional GPU capabilities, such as leveraging the threads to achieve higher performance. 
Important Optix modules that Mochi uses are RayGen, Intersection, and AnyHit. 
{\em RayGen} creates the ray with the user-specified parameters and proceeds to call for BVH traversal and intersection testing. 
For user-defined geometries, {\em Intersection} provides a custom intersection test for ray-primitive intersections. 
The AnyHit program allows the user to record information about the intersection and choose whether to continue or terminate the BVH traversal.

\subsection{Related Work on RT for non-RT applications}
In a mesh consisting of tetrahedrons, Wald et al.\cite{wald} represent a query point as a ray to identify the tetrahedron containing the point.
To perform graph drawing, Zellman et al. \cite{zellman} draw spheres around the points in the dataset, shoot an infinitesimal-length ray to identify the nearest neighbors, and compute  forces exerted by point masses on each other.
Evangelou et al. [\citenum{evangelou}] use a reduction similar to the previous work to perform photon mapping.
To further optimize the same reduction in the context of nearest neighbor applications, Zhu [\citenum{rtnn}] proposed query reordering through morton codes to reduce thread divergence.
They also build several BVHs by partitioning the dataset according to the density of their neighborhood.
Vani et al. [\citenum{dbscan}] accelerate neighbor searches in a clustering algorithm called DBSCAN and form clusters efficiently.
Vani et al. [\citenum{trueknn}] further improve the same reduction by identifying tighter neighborhoods around the query points and iteratively increasing it to find the required number of nearest neighbors.

\subsection{Ray Tracing for Collision Detection}
Both ray tracing and collision detection only care about the spatial positions of the objects in a 3D worldly space.
Because of this common view, they use similar acceleration structures, such as BVH, to efficiently cull the non-object space.
Moreover, {\em intersection} in ray tracing and {\em collision} in collision detection have similar connotations due to the way discrete collision detection is handled.
Between the discrete events when the static frames are picked from an observation of objects over a time interval, the particles would have moved more than just touching, which translates to a spatial intersection of objects. 
Even though collision detection on a static frame closely resembles ray tracing, the scene changes with the position of the camera in ray tracing.
But in collision detection, there is no point of reference; however, the objects themselves can move.
Both ray tracing and collision detection use similar build techniques, yet their goal is different.
The only work that uses RT cores to perform CD, and we are aware of, is an Nvidia white paper \cite{nvidia_white}.
They use rays in the direction of the line of sight to determine the collisions,  which means if a collision is not visible from the camera position, it is not detected.
Weller et al. \cite{weller2013new} also highlight similarities and differences between RT and CD.

\section{Design}
\label{sec:design}
In this section, we start with  porting the existing RT-NNS reduction to solve collision detection and then show it is not a complete solution.
We introduce ideas of Mochi for CD and show how Mochi handles collisions when spheres are of different radii and later on show how this can be generalized to a bounding box with any object that can be described with a succinct mathematical equation.
Finally, we show how Mochi handles the deformable body collisions where the scene is multiple frames of a triangulated mesh.

\subsection{Collision Detection for Uniform Radius Spheres}
\label{sec:uniformradius}
Consider a simple static scene of $n$ particles, where each particle is a sphere of radius $r$ with a non-zero mass and a velocity.
Existing techniques build BVH on the bounding volumes of all the particles in the scene; the bounding volumes could be the particles themselves since a sphere is a simple geometry.
To find the particles that collide with a particle $n_i$, they traverse the BVH by choosing the path that keeps overlapping with $n_i$, and the leaf nodes of this traversal are the colliding particles. 

Doing the same should have been straightforward with the RT-cores as they, too, build the BVH.
To find if two particles are colliding, all we have to do is check if their BVs are overlapping.
Except that, we can not access the BVH built by the RT cores.
The only way to traverse the BVH or access a part of it is by shooting rays from several points in the scene space, hoping they hit the objects.
The existing reduction of Nearest Neighbor Search(RT-NNS) does exactly that to make use of the RT cores.
To find neighbors of a query point $q$ that are at a distance of $r$ from the data points, it first defines spheres of radius $r$ around all the data points. 
Next, it shoots an infinitesimal-length ray with the query point as the ray origin.
All the spheres that this ray intersects are the potential nearest neighbors. 
Ordering these intersections and selecting the top $k$ gives us the $k$ nearest neighbors of $q$.
This can be simply extended to identify the collisions among uniform radii spheres.

\begin{figure}[h]
  \centering
    \includegraphics[width=0.6\linewidth]{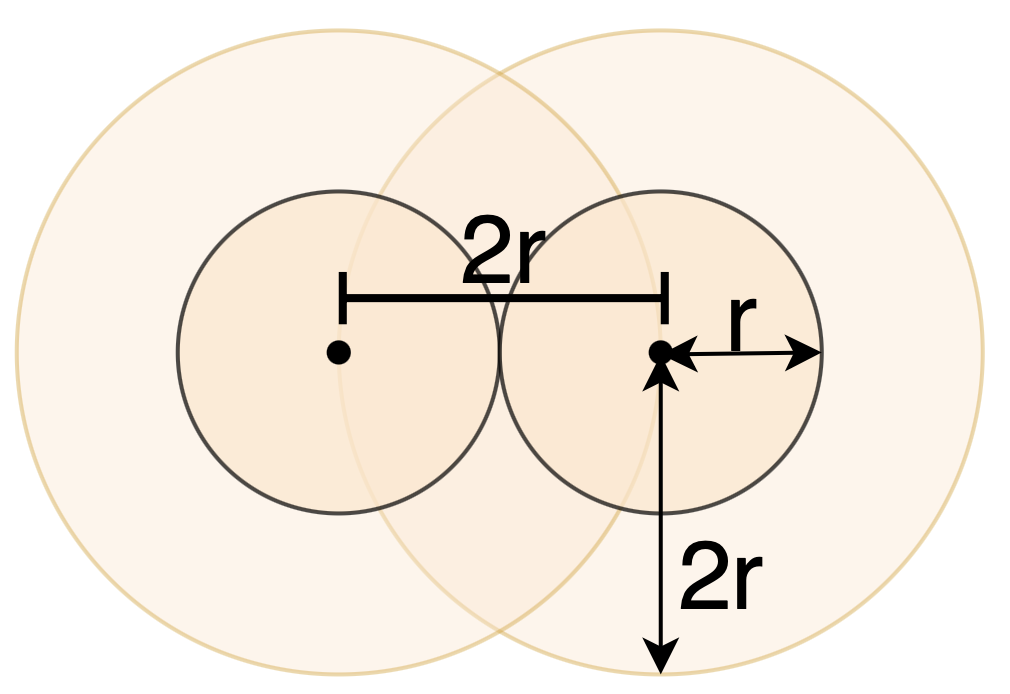}
  \caption{Collision detection on uniform radius spheres.}
  \label{fig:uniformspheres}
\end{figure}

A collision at a given point in time can be an overlap when we consider it in between the discrete time steps.
Hence, at the time of detecting a collision, the distance between the center of colliding spheres could be at most $2r$ and not exactly $2r$.
Given a static frame, finding spheres that collide with $n_i$ means finding spheres whose centers are at a distance of $\le 2r$ from the center of $n_i$.
Essentially, given a point $q$, the goal is to find all the neighboring points that are a distance of $\le 2r$ from $q$, where the points are centers of the spheres.
We reduce the problem of CD to RT-NNS.
For the infinitesimal ray from the RT-NNS reduction to capture these collisions, the colliding sphere would have to contain the ray origin. 
So, Mochi models each sphere with a representative sphere of radius $2r$ as shown in figure \ref{fig:uniformspheres}.
After the BVH is built on all representative spheres, it shoots an infinitesimal ray from the center of each sphere. 
As the ray traverses through the BVH, the spheres that the ray hits are reported by anyHit primitive in no particular order.
In the context of collision detection, a hit simply corresponds to a sphere that collides with the sphere at whose center the ray originated.

\vspace{0.2cm}
\begin{reduction} \label{red:1}
In a scene of spheres of radius $r$, to detect all the collisions, Mochi builds a representative sphere of radius $2r$ around every sphere with the center remaining the same and follows the RT-NNS reduction.
\end{reduction}

\subsection{Non-uniform Radii Spheres}
\label{sec:nonuniform}

In this section, we consider the particles with different radii. 
The scene is a set of $n$ particles, where each particle $n_i$ is a sphere of radius $r_i$.
For any two particles $n_i$ and $n_j$ with radii $r_i$ and $r_j$, $r_i$ is not necessarily equal to $r_j$.
If we follow the same reduction as in section \ref{sec:uniformradius} where we use a representative sphere of radius $2r_i$ for the sphere $n_i$, we capture all the intersections that are at a distance of $2r_i$ from the center of $n_i$.
However, the distance between the centers of two colliding spheres $n_i$ and $n_j$ is at most $r_i+r_j$.
Trying to make use of RT-NNS causes the reported set of collisions to contain several false positives and negatives. 
If $r_i<r_j$, several actual collisions are missed.
If $r_i>r_j$, several collisions are reported when they might not have happened.
Another simple solution is to use representative spheres of radius $r_{max}: r_{max}\ge r_i \ \forall n_i $, where $r_{max}$ is the maximum radius among all the radii in the scene.
This solution ensures correctness; however, it incurs large runtimes as pointed out by a previous work \cite{trueknn}.

A previous work on performing CD using ray tracing \cite{nvidia_white} shoots a ray along the line-of-sight with the position of the camera since they are only concerned with rending the scene for visual graphics.
This ray is not of infinitesimal length, but it clearly misses all the other collisions that are not in the line of sight.
To make up for this, let us say we shoot rays in multiple directions; there can always be an object between the rays unless the rays are dense enough not to miss the smallest sphere.
To completely capture all the collisions, we need to find a way to closely follow along the object's periphery since that is where the collisions happen.
Reduction~\ref{red:1} shoots rays from the center when the collisions happen far away from the center, given a reasonable time step.
Instead, we propose to shoot rays of finite length from the corners of the bounding box than rays of infinitesimal length from the center of the object.

RT-cores support rendering triangles, b-spline curves, and all the other objects through AABBs.
Hence, the particles, which are spheres, will be bounded by a rectangle and cuboid in $2D$ and $3D$, respectively.
To closely trace the periphery of the bounding box of a sphere, Mochi launches rays along the edges of this box. 
A ray starts at one vertex and ends at the other vertex of an edge of the box, so the ray length is equal to the edge length that it is tracing.
The direction of the ray could be either clockwise or anti-clockwise; Mochi chooses the directions such that as many rays could be launched from a single vertex of the bounding box. 
The number of rays launched per particle equals the number of edges of the AABB, which are $4$ and $12$ in $2D$ and $3D$, respectively.
As the rays are scouting the periphery of the bounding volume, at least one of the colliding objects' edge rays will report the collision. 
This suffices since collision is commutative.

\begin{figure}[h]
  \centering
    \includegraphics[width=\linewidth]{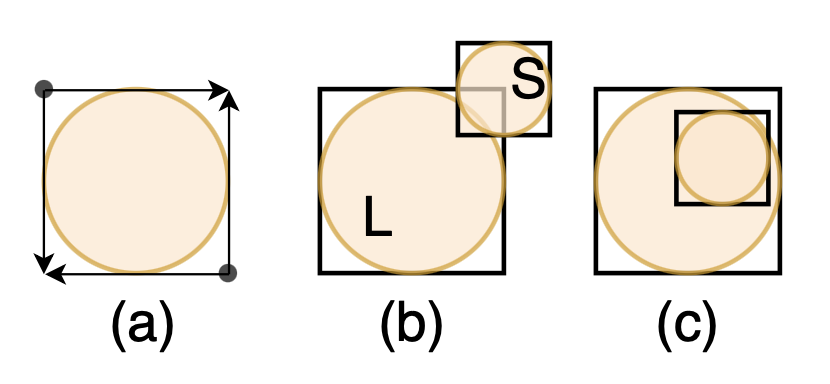}
    \vspace*{-1cm}
  \caption{Collision detection on non-uniform radius spheres: a)tracing the AABB of a sphere by shooting rays along the edges of AABB b)AABB of two colliding spheres overlap but not completely c)AABB of one of the two colliding spheres is completely contained inside the other.}
  \label{fig:nonuniform}
\end{figure}

Consider two moving spheres $n_i$ and $n_j$ with $r_i$ and $r_j$, respectively, that are colliding.
As shown in figure \ref{fig:nonuniform}, we have the following two cases.
\begin{description}
\item[Case 1:] The bounding volumes overlap with each other but are not completely nested.
The edge rays of both the AABBs intersect the other's AABB. 
Both the objects report the broad phase collision.\\
\item[Case 2:] The smaller of the AABBs is completely nested in the large AABB. 
The edge rays of larger AABB will miss the collision; however, the edge rays of smaller AABB report this intersection.
\end{description}

Once the rays find the intersections with the AABBs, there will be some false positives since Mochi only looked at intersections between AABBs and not the spheres themselves up until now.
With these intersections, Mochi moves on to the narrow phase of collision detection.
For an intersection reported by the RT-cores, Mochi computes the distance between the centers of the spheres whose AABBs intersected and marks it as a collision only if this distance is $\le r_i+r_j$

In the above process, the ray-AABB intersection test is hardware accelerated by RT-cores, while the computation required for the narrow phase is processed by a shader core.
Each of the edge rays traverses the BVH and reports the intersections independently from the other rays.
There could be a data race while marking the collision as multiple edge rays of a sphere could have detected the same AABB for collision.
However, since the threads are simply marking, the output does not change regardless of the order in which threads write the collision data.

\vspace{0.2cm}
\begin{reduction} \label{red:2}
In a scene of spheres with non-uniform radius, to detect all the collisions, Mochi shoots rays along the edges of the AABB bounding the sphere.
\end{reduction}

\subsection{Generic Object Collision Detection}
Exact collision detection means we need to be able to accurately find out if two objects are intersecting.
Despite the fact that the RT-cores only provide an interface to check if a ray is intersecting an object, Mochi performs an object-object intersection test by extending the reduction \ref{red:2}.
In the broad phase, Mochi only identifies the AABB intersections.
While some false positives are reported, all the other objects that can not potentially cause a collision are removed.
It is the narrow phase that Mochi generalizes to capture generic object-object collisions.
The algorithm \ref{alg:1} details this entire procedure.

\begin{algorithm}
\caption{Mochi object collision detection}\label{alg:1}
\begin{algorithmic}[1]
\State $\forall\ 1\le i\le n: AABB[i] = bounds(object[n_i])$  \label{line:1} 
\SubAlgorithm{Launch Rays:} \label{line:2} 
\State $orgins \gets computeAABBVertices()$ \label{line:3}
\State $lengths \gets computeAABBDimensions()$  \label{line:4}
\State $dirs \gets \{(\pm 1,0,0),(0,\pm 1,0),(0,0,\pm 1)\}$   \label{line:5}
\State $\textbf{for}\ i=1\ to\ n\ \textbf{do}$  \label{line:6}
\State \hskip0.5em $\textbf{for}\ k=1\ to\ 12\ \textbf{do}$ \label{line:7}
\State \hskip1em $traceRay(origins[n_i*k], dirs[k], lengths[n_i*k])$    \label{line:8}
\EndSubAlgorithm
\SubAlgorithm{BVH traversal by every ray:}   \label{line:9}
\IF {an intersection is reported}   \Comment{Broad phase}\label{line:10} 
    \State $n_i \gets allObjects[getRayID()/12]$   \Comment{Narrow phase}\label{line:11}
    \State $n_j \gets getIntersectedObject()$   \label{line:12}
    \State $l \gets getRayTMax()$   \label{line:13}
    \State $P \gets ray.origin + l*ray.direction$   \label{line:14}
    \IF {$isInside(n_i,P)\ and \newline \hspace*{1cm} isInside(n_j,P)$}  \label{line:15} 
        \State $markCollision(n_i,n_j)$   \label{line:16}   
    \ENDIF 
\ENDIF 
\EndSubAlgorithm
\end{algorithmic}
\end{algorithm}

As long as a mathematical equation that describes the object and a function(`isInside') that determines if a point is inside the object are given as input to Mochi, it takes only a step extra for Mochi to find exact collisions from the reported AABB intersections.
In algorithm \ref{alg:1}, line \ref{line:1} defines a tightly fit AABB by choosing the maximum and the minimum (x,y,z) coordinates based on the object definition.
At this point, Mochi makes a call to RT cores via Optix API to construct the BVH.
Lines \ref{line:2}-\ref{line:8} launch rays to trace the edges of AABB of every object as explained in section \ref{sec:nonuniform}.
As these rays traverse the BVH and find the AABBs of potential collisions, Mochi executes the narrow phase in lines \ref{line:11} through \ref{line:16}.
The line \ref{line:13} gets the length of the ray from the source to the point of intersection.
In line \ref{line:14}, the point of intersection between the ray and object inside hit AABB is computed.
In line \ref{line:15}, Mochi determines if this point of intersection lies inside both the hit object and the object the ray corresponds to.

In algorithm \ref{alg:1}, the broad phase is entirely handled by the RTX hardware. 
RT cores report only the candidate collisions to Mochi, thus accelerating the entire CD pipeline.
In the narrow phase, it is crucial that Mochi performs a minimal number of object-object intersection tests and instructions per intersection test. 
The instructions of the narrow phase are executed by shader cores, similar to any other GPU instruction.
If we do not set the scene efficiently and provide an optimal `inside' function, the narrow phase could offset the performance gain achieved through hardware acceleration by the broad phase.

\vspace{0.2cm}
\begin{reduction} \label{red:3}
To detect all the collisions in a scene of objects given their mathematical definitions and a function(isinside) to check if a point lies inside the object, Mochi shoots rays along the edges of the AABB bounding the sphere and calls isinside() at the point of intersection.  
\end{reduction}

\subsection{Deformable body collisions}
A deformable body differs from a rigid body in that the relative positions of two points that are part of the object might change with time.
This means parts of the body may also collide with themselves. 
We can no longer treat a deformable body as an atomic object to detect all collisions in the scene. 
In literature, deformable objects are broken down into polygons, usually triangles.
Moreover, it is difficult to describe such complex objects by an equation.

The scene comprises a triangle mesh with all the objects in the frame triangulated.
A set of vertices and triangles built on these vertices form the triangle mesh.
We must find all pairs of triangles that intersect to find the parts of the object that collide with itself or other objects.
Fortunately, the RT cores are built to support triangles as a data type.
However, the underlying hardware interface provides functionality only for a ray-triangle intersection test but not a triangle-triangle intersection test.

\subsubsection{Triangle-Triangle Collision}
Finding if two triangles $T_i, T_j$ collide is computationally intensive.
Any of the three edges of $T_i$ can intersect the face of the other triangle $T_j$ and vice-versa.
There is extensive research on making these tests faster \cite{moller1997fast, tropp2006fast}.
To perform fewer tests, BVHs are also used to index triangles of a triangle mesh.
A straightforward solution for performing CD on triangle meshes using RT cores is to use reduction \ref{red:2} by  bounding triangles with AABBs.
There are multiple downsides to this approach.
First, RT cores support triangles as a datatype, and the fastest test they can perform is a ray-triangle intersection. 
To use a ray-AABB intersection test over a ray-triangle intersection test is an ineffective use of RT core capabilities.
Second, a triangle has no volume, which means most of the corresponding AABB is empty.
This, in turn, increases the runtime of the narrow phase since the hit in a broad phase does not have much significance. 
A triangle has three edges, whereas an AABB has 12 edges.
{\em Tracing the triangle along its edges instead of tracing its bounding box} would be a more efficient solution.
However, this approach does not capture intersections when the ray along the edge is coplanar to the target triangle.
In terms of rending a scene, rendering a triangle through which a ray is passing does not make sense.
The triangle should have some depth to have any optical effect.
But then, it will not be a triangle anymore and should be bounded by an AABB.
RT cores' behavior is undefined in the case of a coplanar ray-triangle intersection test, most likely not being able to detect it.

\begin{figure}[h]
  \centering
    \includegraphics[width=0.9\linewidth]{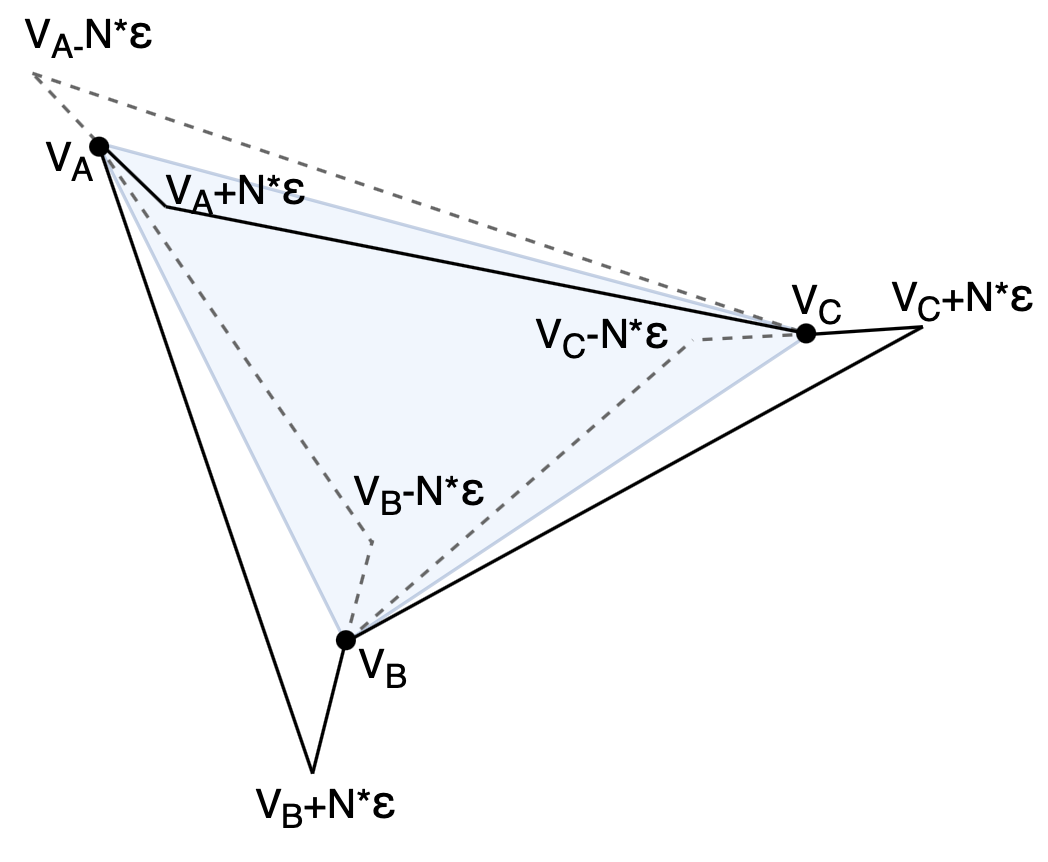}
  \caption{triangles}
  \label{fig:triangle}
\end{figure}

Coplanar ray-triangle intersection means that the initial point of contact between the ray and the triangle lies on one of the sides/edges of the triangle.
Observe that the ray passes through a plane orthogonal to the plane containing the triangle.
If we place three orthogonal planes such that each orthogonal plane contains one side of the triangle, then any coplanar ray intersecting the triangle will intersect one of these planes.
To leverage the native support of RT cores for triangles, we propose building triangles instead of planes.
Figure \ref{fig:triangle} describes our approach.
Along each edge of the triangle, we construct an auxiliary triangle such that 1)the plane of the auxiliary triangle is orthogonal to the plane of the original triangle, 2) one of the vertices of the auxiliary triangle is one of the endpoints of the original triangle edge, and 3) a median of the auxiliary from this chosen vertex is the original triangle edge.
Let $T(V_A,V_B,V_C)$ represent the original triangle with $V_A,V_B,V_C$ as its vertices and $N$ as the direction of normal to the triangle plane.
Then the three auxiliary triangles can be represented as $T(V_A, V_B+(\epsilon * N), V_B-(\epsilon * N))$, $T(V_B, V_C+(\epsilon * N), V_C-(\epsilon * N))$, and $T(V_C, V_A+(\epsilon * N), V_A-(\epsilon * N))$ where $\epsilon$ is an infinitesimally small positive number.

Mochi constructs three auxiliary triangles for every triangle and indexes all four of them using the BVH.
As in reduction \ref{red:3}, it shoots rays along the triangle edges but only for the original triangle.
Since the triangle has three edges, Mochi shoots three rays, one along each edge, to capture triangle-triangle collisions.
The broad phase ends when an edge ray intersects a triangle, whether it is the original or the auxiliary.
In the narrow phase, Mochi first determines the type of triangle that is intersected.
An intersection with the original triangle implies a triangle-triangle intersection, while an intersection with an auxiliary triangle implies only a potential collision.
To determine if the ray-auxiliary-triangle intersection leads to actual collision, Mochi computes the point of intersection and checks if it falls on the particular median, which is also the edge of the original triangle.
Mochi's triangle-triangle intersection test in the narrow phase is much simpler than the tests popular libraries run on their potential triangle colliding pairs.

The edge rays of a triangle $T_1$ will report all collisions unless the other triangle $T_2$ is intersecting only the face of $T_1$, in which case the edge rays of the intersecting triangle $T_2$ will report the collision.
Mochi replies on the commutativity of collision to detect all pairs of triangle collisions.
At least one of the 6 edge rays involved in detecting a collision between a pair of triangles $T_1, T_2$  will report the intersection.

\paragraph*{self-collisions} Because Mochi considers all pairs of triangles for collisions regardless of which object in the scene the triangle belongs to, self-collisions will also be reported and decoded as long as the triangles are marked appropriately to which object they belong. If the objects are not marked, a simple connected component test similar to Oibvh\cite{oibvh} can be performed before or after CD.

\vspace{0.2cm}
\begin{reduction} \label{red:4}
To detect all the collisions, including self-collisions, on a triangle mesh, Mochi first constructs 3 auxiliary triangles along its edges and indexes all the triangles. Then, Mochi shoots rays along the edges of the original triangle to capture all collisions.
\end{reduction}

\subsection{Moving objects}
In the case of dynamic scenes, the positions of the objects change and are described using multiple frames as they deform through space and time.
Because the objects are moving, the BVH built for one frame is not accurate during the next frame.
Some works make use of temporal and spatial coherency to avoid rebuilding the BVH for every frame.
For example, remembering the list of nodes visited during the first BVH traversal and rechecking them during the second frame's BVH traversal. 
Refitting the BVH, where parts of the tree are modified to reflect current objects' positions, is another popular strategy. 
However, these techniques are only helpful if most of the objects in the scene are static.
So Mochi implements both refit and rebuilding BVH every frame and thoroughly evaluates the costs in section \ref{sec:evaluation}.

\section{Implementation} 
\label{sec:implementation}
In the background section \ref{sec:background}, we describe the accessible RT cores functionalities.
To implement our reductions, Mochi uses AnyHit primitive to capture collisions after issuing the rays using Raygen.
Any operation involving triangles is comparably faster than those involving AABBs.
While hardware acceleration is the main reason, we also note that triangle geometry uses the structure of arrays format, unlike the other geometries that need to use the array of structures.
Mochi offloads all the rays at once to RT cores without sorting them.
We did not add Morton code optimization even though it was proven helpful in the case of nearest neighbor search.
Since our preliminary experiments did not show an improvement in collision detection times.

Our reductions do not involve any significant computations.
The only bottleneck we faced was that of memory.
Mochi uses a bit array to mark the collisions.
1 byte holds collision information for eight pairs of objects.
So, Mochi is limited to processing only the object pairs that fit in the memory at a time.
Mochi can handle millions of objects if we do not need to store the collisions.
Moreover, our experiments showed that Oibvh faces memory issues faster than Mochi.

\paragraph*{End-to-end particle simulation}
Using reduction \ref{red:2}, Mochi simulates end-to-end particle collision detection.
It supports particles with different radii and attributes such as mass and velocity.
It also handles multi-particle collisions.
After a chosen timestep, it detects all the collisions, calculates the change in velocity and positions, and updates them for the next time step.
Mochi rebuilds the BVH to account for the changes in the scene.

\section{Evaluation}
\label{sec:evaluation}

\begin{description}[leftmargin=*]
\item[Experimental Setup]
We used NVIDIA GeForce RTX 2060 GPU with 6GB memory, 1920 CUDA cores, and 30 first-generation RT cores.
Mochi uses Optix Wrapper Library \cite{owl} to interact with RT and shader cores of the GPU.

\item[Baselines]
We compare Mochi with OIBVH \cite{oibvh}, the state-of-the-art GPU-based collision detection library.
OIBVH code takes only a triangle mesh as input, so we modified it to take spheres as input.

\item[Datasets]
We generate sphere positions and radii for particle collision systems following uniformly random and Gaussian distributions.  
We chose the range of parameters, the number of spheres in the scene, and the radius of each sphere based on literature survey \cite{breuer2015modeling,dehnen2011n} and hardware constraints.
For triangle meshes, we chose the same datasets as OIBVH, which are described in Table \ref{table:datasets}.
All the dataset descriptive videos are available on MCCD \cite{TMT10-GMOD} website.
These datasets are in ply format, and we use rply library to read them.

\begin{table}[h!]
\centering
\begin{tabular}{ |c|c|c|c|l| } 
 \hline
 Dataset & Vertices & Triangles & Frames & \multicolumn{1}{p{1.2cm}|}{avg collisions per frame}\\ \hline
 Funnel & 9450 & 18484 & 500 & 0.83\\ 
 Flamenco & 25686 & 49218 & 706 & 1364.97\\
 Cloth-ball & 46598 & 92230 & 94 & 1.62\\ 
 N-body & 73960 & 146480 & 76 & 3899.92\\ 
 \hline
\end{tabular}
\caption{Triangle mesh datasets}
\label{table:datasets}
\end{table}

\item [Performance Evaluation]
We verified the correctness of Mochi by generating the ground truth from a program that considers every object pair for collision.
We further verify the collision pairs detected by Mochi with OIBVH as well.
Every performance number reported in the following experiments is averaged over five runs. 
For all particle experiments, we plot performance numbers until Oibvh runs without any memory issues.
Oibvh runs out of memory faster than Mochi since it stores all the broad phase collisions and then forwards them to a narrow phase, unlike Mochi, which performs a narrow phase check as soon as a broad phase collision is detected.

\end{description}

\subsection{Particle system collisions}
We wanted to evaluate end-to-end particle system simulation by detecting collisions and updating their positions and velocities at every time step. 
However, after the literature study, we found that CD is performed at a particular time step, i.e., only per frame performance is evaluated.
So, we first evaluate Mochi's performance as the number of particles in the system increases by keeping the density of the scene constant.
Next, we vary the density by increasing the radius of the particles.

\subsubsection{Changing the number of spheres}
\begin{figure}[h]
  \centering
    \includegraphics[width=0.8\linewidth]{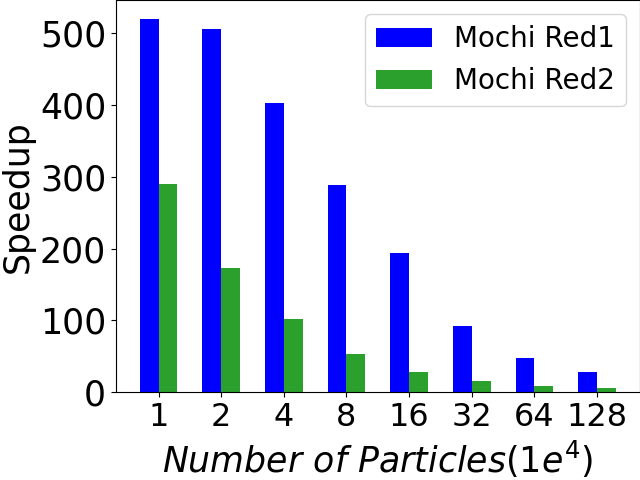}
  \caption{CD time speedup of Mochi reduction \ref{red:1} (Red1) and \ref{red:2} (Red2) over Oibvh as the number of particles increase from 10K to 1.28M. radius = 0.001}
  \label{fig:red1_2_n}
\end{figure}

\begin{figure}[h]
  \centering
    \includegraphics[width=0.8\linewidth]{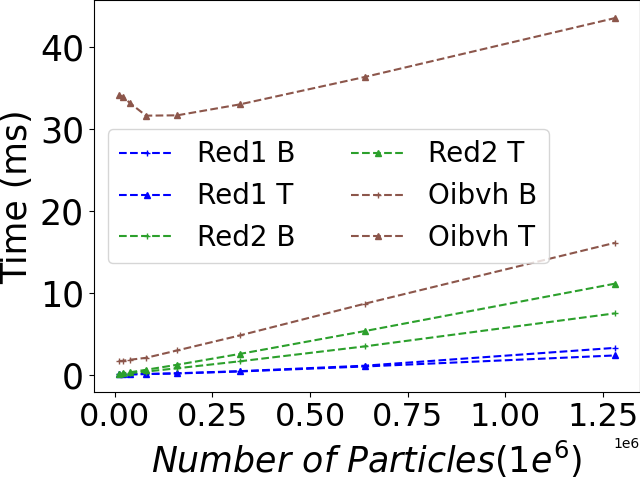}
  \caption{comparison of total CD time (T) and broad phase time (B) of Mochi reduction \ref{red:1} (Red1) and \ref{red:2} (Red 2) over Oibvh as the number of particles increases from 10K to 1.28M. Radius = 0.001. Red1 B and Red1 T stand for Mochi's reduction \ref{red:1} total cd time and broad phase time. Similarly, other abbreviations.}
  \label{fig:n1}
\end{figure}

Figure \ref{fig:red1_2_n} shows the speedups of Mochi reduction \ref{red:1} and reduction \ref{red:2} against the baseline as the number of particles in the scene increases from 10K to 1.28M.
We see that the speedups of both reductions decrease as the number of particles increases.
Speedups range from 520x-28x and 289x-6x for reduction \ref{red:1} and \ref{red:2} respectively.
We expected this decrease in speedups to be due to a memory contention issue.
So, we recreated this experiment on a Titan V GPU with 12GB memory, which is twice RTX 2060.
However, it also displays almost identical behavior.
So, we decoupled the performance of Mochi's reductions from Oibvh in figure \ref{fig:n1}.
It displays the total collision detection time and broad phase time for the same runs as in figure \ref{fig:red1_2_n}.
While Mochi reductions spend more time in the broad phase with a negligible amount of time in the narrow phase, Oibvh spends significantly less time in the broad phase compared to its narrow phase.
But with the increase in the number of particles, both Mochi and Oibvh spend increasingly more time in the broad phase.
The time spent in the broad phase is further proportional to the number of threads each implementation spawns.
Even though RTX architecture as a whole uses both shader and RT cores, the number of RT cores in RTX 2060 is much smaller compared to the number of shader cores, thus limiting Mochi's performance as the number of particles increases.

Figure \ref{fig:red1_2_n} also shows that speedups of reduction \ref{red:1} are higher than that of reduction \ref{red:2}. 
While reduction \ref{red:1} uses eight times larger bounding volumes, reduction \ref{red:2} uses 12 times more number of rays (mapped to threads).
The number of collisions is only a few hundred.
So, we note that using a higher number of threads significantly affects runtime regardless of collisions in the broad phase.
The traversal of BVH itself by several threads is time-consuming.

\subsubsection{Changing the size(radius) of sphere}
\begin{figure}[h]
  \centering
    \includegraphics[width=0.8\linewidth]{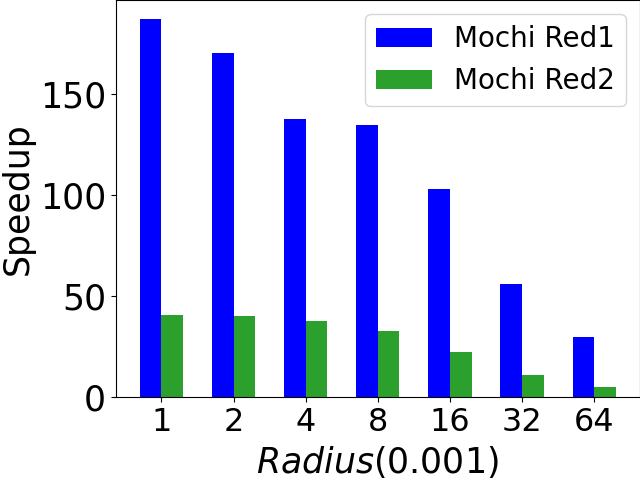}
  \caption{CD time speedup of Mochi reduction \ref{red:1} (Red1) and \ref{red:2} (Red2) over Oibvh as the radius of particles increases from 0.001 to 0.064, number of particles in the scene = 160,000}
  \label{fig:red1_2_r}
\end{figure}

\begin{figure}[h]
  \centering
    \includegraphics[width=0.8\linewidth]{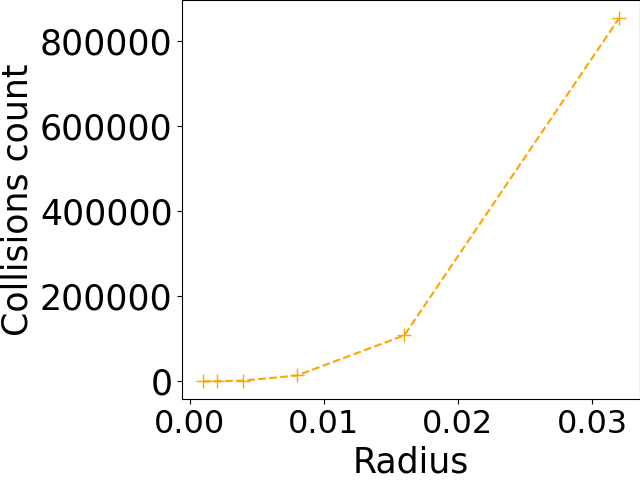}
  \caption{Number of particle collisions in the scene as the radius increases from 0.001 to 0.064, number of particles in the scene = = 160K}
  \label{fig:red1_2_r_count}
\end{figure}

Figure \ref{fig:red1_2_r} shows the speedups of Mochi reduction \ref{red:1} and reduction \ref{red:2} against the baseline as the size of all the spheres increases from 0.001 to 0.064.
Note that all the spheres in each scene have the same radius.
The radius is the only parameter changing from scene to scene, with the centers of spheres remaining the same.
We observe that speedups decrease with an increase in radius.
Speedups range from 187x-30x and 41x-5x for reduction \ref{red:1} and \ref{red:2}, respectively.
For reduction \ref{red:1}, an increase in radius means larger number of AABB intersections than reduction \ref{red:2}, since AABB in reduction \ref{red:1} is eight times larger than AABB in reduction \ref{red:2}.
For reduction \ref{red:2}, an increase in radius by a factor of 2 means longer rays with twice the length.
While reduction \ref{red:1} is still faster than reduction \ref{red:2}, the workload is increasing at a higher rate for reduction \ref{red:1} than reduction \ref{red:2} with an increase in radius.
To support this observation, we plot the number of collisions at the end of a narrow phase in figure \ref{fig:red1_2_r_count}. 
We see that the number of collisions increases by a factor of 10 with an increase in radius by a factor of 2.
Since reduction \ref{red:1} is using a larger radius than the actual radius of the sphere, it is capturing an unnecessarily higher number of AABB intersections than reduction \ref{red:2} at a given radius.
Moreover, reduction \ref{red:2} has a steady speedup, implying that its workload is increasing at the same rate as Oibvh with an increase in r, as they are both same-sized AABBs.

\begin{figure}[h]
  \centering
    \includegraphics[width=0.8\linewidth]{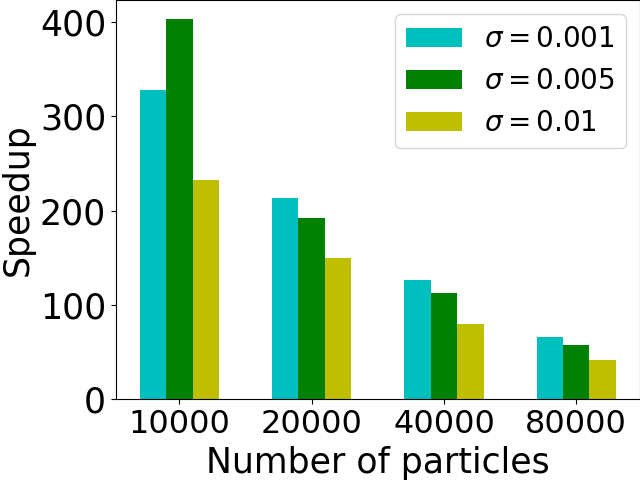}
  \caption{CD time of Mochi reduction \ref{red:2} over Oibvh with an increase in the number of particles and the variation in the radius of the particles. mean radius=0.004}
  \label{fig:diff_r}
\end{figure}

So far, we evaluated the reduction \ref{red:2} on scenes with all the spheres having the same radius.
But the power of reduction \ref{red:2} is that it can support spheres with different radii, unlike the reduction \ref{red:1}.
In figure \ref{fig:diff_r}, we compare the reduction \ref{red:2} of Mochi with Oibvh as we increase the number of particles and the variation in the radius of the particles in the scene.
On the X-axis, we increase the number of particles from 10K to 80K.
For each set number of particles, we vary the standard deviation as 0.001, 0.005, and 0.01, while keeping the mean radius at 0.004.
Mochi's reduction \ref{red:2} speedups are not much different from the previous experiments.
As the radius of the particles increases in the scene or between the scenes, the speedups decrease.
The speedups also decrease when the number of particles increases.
RT implementations are greatly effected by the size and the number of AABBs. 

\subsubsection{BVH build times}
\begin{figure}[h]
  \centering
    \includegraphics[width=0.8\linewidth]{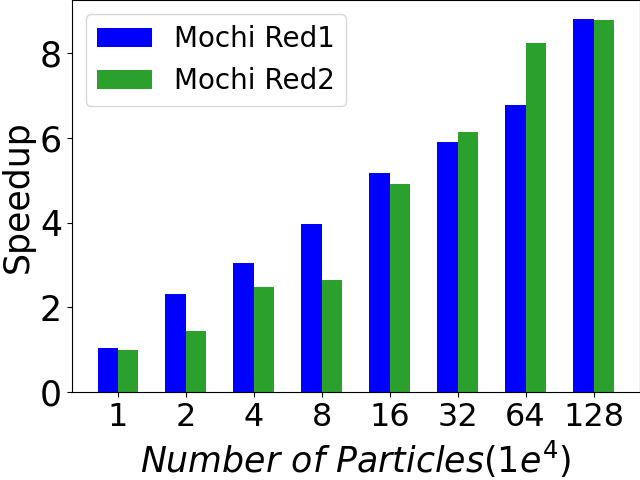}
  \caption{Build time speedup of Mochi reduction \ref{red:1} (Red1) and \ref{red:2} (Red2) over Oibvh as the number of particles increases from 10K to 1.28M. radius = 0.001}
  \label{fig:red1_2_n_build}
\end{figure}

Figure \ref{fig:red1_2_n_build} shows the speedup of Mochi build time over Oibvh build time as the number of particles increases from 10K to 1.28M.
Speedups range from 9x-1x both  reductions \ref{red:1} and \ref{red:2}.
These numbers are from the same experiment corresponding to figure \ref{fig:red1_2_n}.
The build times of all implementations increase with the increase in n. 
Mochi's reductions build BVH faster than Oibvh's since Mochi uses RT cores.
Moreover, Oibvh spends time sorting the objects before building the BVH.
Though the build time speedups are not as high as CD times, the interesting trend is that RT cores are increasingly beneficial for building BVH when there are a higher number of objects to index.

\begin{figure}[h]
  \centering
    \includegraphics[width=0.8\linewidth]{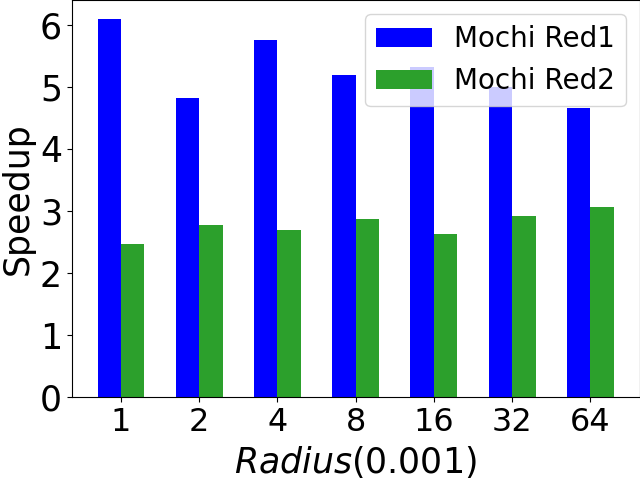}
  \caption{CD time speedup of Mochi reduction \ref{red:1} (Red1) and \ref{red:2} (Red2) over Oibvh as the radius of particles increases from 0.001 to 0.064, number of particles in the scene = = 160K}
  \label{fig:red1_2_r_build}
\end{figure}

Figure \ref{fig:red1_2_r_build} shows the speedup of Mochi build time over Oibvh build time as the radius of particles increases from 0.001 to 0.064.
These numbers are from the same experiment corresponding to figure \ref{fig:red1_2_r}.
We do not observe any trend with a change in radius.
Computing AABBs does not depend on the radius. 
The bounds of the box are simply chosen by applying maximum and minimum functions on the object's coordinates.

\subsection{CD on triangle meshes}

\begin{table*}
\centering
\begin{tabular}{ |c|l|l|l|l|l|l|l|l|c| } 
 \hline
  & \multicolumn{3}{|c|}{\bfseries Mochi Refit} & \multicolumn{3}{|l|}{\bfseries Mochi Rebuild} & \multicolumn{2}{|c|}{\bfseries Oibvh} &  \multicolumn{1}{|c|}{}\\ \hline
 Dataset & \multicolumn{1}{|p{0.8cm}|}{Initial build time} & \multicolumn{1}{p{1.1cm}|}{avg refit time per frame} & \multicolumn{1}{p{1cm}|}{avg CD time per frame} & \multicolumn{1}{|p{0.8cm}|}{Initial build time} & \multicolumn{1}{p{1.2cm}|}{avg rebuild time per frame} & \multicolumn{1}{p{1cm}|}{avg CD time per frame} & \multicolumn{1}{|p{1.1cm}|}{avg build time per frame} & \multicolumn{1}{p{1.3cm}|}{avg CD (broad phase) time per frame} & \multicolumn{1}{|p{1cm}|}{Speedup}\\ \hline
 Funnel &  1.93 & 0.69 & 0.20 &  2.16 & 1.79 & 0.14 & 2.26 & 1.82 (1.56)& 13 \\ 
 Flamenco & 2.78 & 1.63 & 0.36 & 2.16 & 2.87 & 0.24 &  6.62 & 6.92 (2.94) & 28.83 \\ 
 Cloth-ball & 4.23 & 2.31 & 0.59 & 2.72 & 4.07 & 0.35 & 7.71 & 1.85 (1.58) & 5.28\\ 
 N-body & 4.61 & 4.00 & 7.83 & 3.04 & 5.45 & 0.39 & 63.55 & 9.68 (3.25) & 24.8 \\ 
 \hline
\end{tabular}
\caption{Overall comparison of Mochi and Oibvh on triangle mesh datasets, all times in milliseconds}
\label{table:overall}
\end{table*}

In table \ref{table:overall}, we compare the reduction \ref{red:4} of Mochi with unmodified Oibvh on several real-world triangle mesh datasets, which are described in table \ref{table:datasets}.
Each dataset describes a moving scene and, hence, has multiple frames.
After each frame, the BVH is usually refitted for the next  frame without building a whole new tree since the refit is faster. 
However, this comprises the quality of BVH to a certain extent and could potentially make traversals longer.
Oibvh doesn't refit but instead makes a fresh build every frame.
They note that since their build times are faster than the previous state-of-the-art, they could bear the costs of  building a new BVH every frame to best preserve BVH quality.
Hence, we implemented both refit and rebuild in Mochi.
Refit means that parts of BVH might change, while rebuild means that the whole BVH will be replaced by a new one.
Our rebuild is slightly different than Oibvh's.
BVH is one of the several accelerating structures in the Optix program run by RT cores.
During the rebuild, only BVH is reconstructed, leaving the others.
And that is also the reason why Mochi's rebuild is slower than its refit, but it is not as slow as the initial build.
Hence, Oibvh's average build time (first sub-column of Oibvh in table \ref{table:overall}) is comparable to the initial build times of Mochi rather than average refit or rebuild times (second sub-columns of Mochi Refit and Rebuild respectively).

The second observation we make regarding Mochi's indexing time is that the initial build time (columns 2 and 5) is affected by whether Mochi refits or rebuilds in consecutive frames.
Mochi's Rebuild initial build time is faster than its Refit initial build time. However, it gets offset when rebuilding BVH during consecutive frames takes more time than refitting.
Since refit might change only the parts of the tree, the RT cores spend more time optimizing the BVH during the initial build.
Thus, refitting BVH takes less time during consecutive frames.
Whereas in the case of rebuild, the initial build times and average rebuild times for later-on frames are not much different.
In fact, later frame rebuild times are higher due to the fact that the scene gets complex as the frames progress.

Whether Mochi refits or rebuilds, its triangle indexing is faster than Oibvh's for all the datasets.
Using reduction \ref{red:4}, Mochi builds 4 times more triangles, including the auxiliaries, and yet, it builds faster than Oibvh without losing on the collision detection performance.
This clearly shows the impact of hardware acceleration leveraged by Mochi.
On a side note, the reason Oibvh spends more time in building BVH is because it computes bitonic ordering, Morton codes, and special AABB node layouts to make the traversal (collision detection) faster.
The final observation regarding build times is that all build times (Mochi's refit and rebuild initial build times and Oibvh's average build time) increase with the increase in the number of triangles.
For the N-body dataset, Oibvh especially takes more build time because this dataset contains several individual objects compared to other datasets.
Oibvh builds a separate BVH for each of the objects, unlike Mochi which builds only one BVH over the entire dataset.

\subsubsection{Collision Detection(CD) times}
The speedup of Mochi's Rebuild average CD time (last sub-column of Mochi Rebuild) over Oibvh's average CD time is noted in the last column of the table \ref{table:overall}.
Mochi is 5x-29x times faster than Oibvh over real-world datasets.
To investigate these speedups, we once again turn to the number of AABB intersections each of these libraries detect and report them in table \ref{table:intersection}.
Surprisingly, Mochi finds more broad phase intersections than Oibvh even though Mochi is faster.
RT cores traverse the BVH and report the colliding triangle pair to the user, Mochi.
Whereas Oibvh chooses the starting point of the traversal and directs the traversal by itself.
Not only in the case of building BVH but also in the case of traversing BVH, we see the impact of hardware acceleration.
The observation that Mochi finds more AABB intersections than Oibvh sheds light on the quality of BVH that RT cores construct.
The BVH built by RT cores is not necessarily more efficient or of higher quality than Oibvh.
It is simply hardware versus software acceleration.

\begin{table}[h!]
\centering
\begin{tabular}{ |c|c|c| } 
 \hline
 Dataset & Mochi & Oibvh \\ \hline
Funnel & 5692.01 & 16169.07 \\
Flamenco & 19016 & 124771.49  \\
Cloth-ball & 28704.6 & 21458.04 \\
N-body & 48597.5 & 20513.09  \\
 \hline
\end{tabular}
 \caption{Broad phase intersections on triangle mesh datasets}
\label{table:intersection}
\end{table}

\begin{figure*}
    \centering
    \begin{minipage}[b]{\textwidth}
        \centering
        \begin{subfigure}[b]{0.24\textwidth}
            \includegraphics[width=\linewidth]{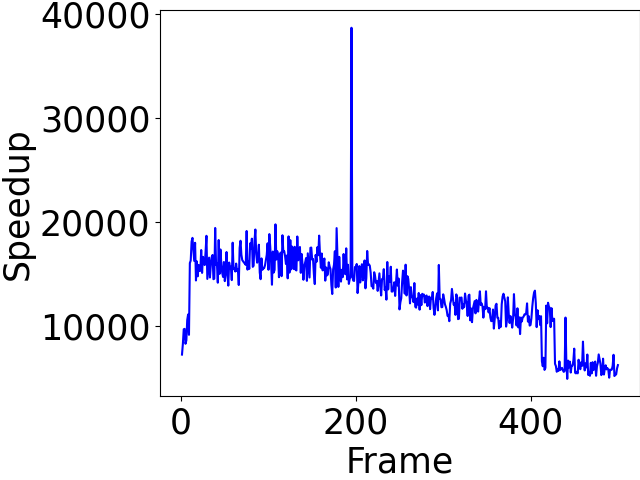}%
            \caption{Funnel}
            \label{fig:5a}
        \end{subfigure}
        \hfill
        \begin{subfigure}[b]{0.24\textwidth}
            \includegraphics[width=\linewidth]{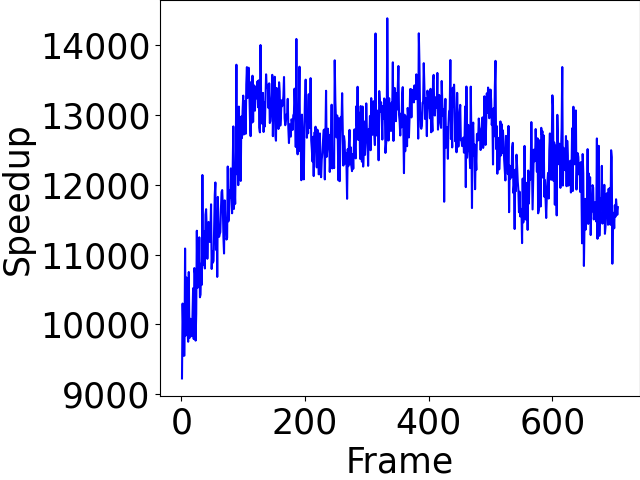}
            \caption{Flamenco}
            \label{fig:5b}
        \end{subfigure} 
        \hfill
        \begin{subfigure}[b]{0.24\textwidth}
            \includegraphics[width=\linewidth]{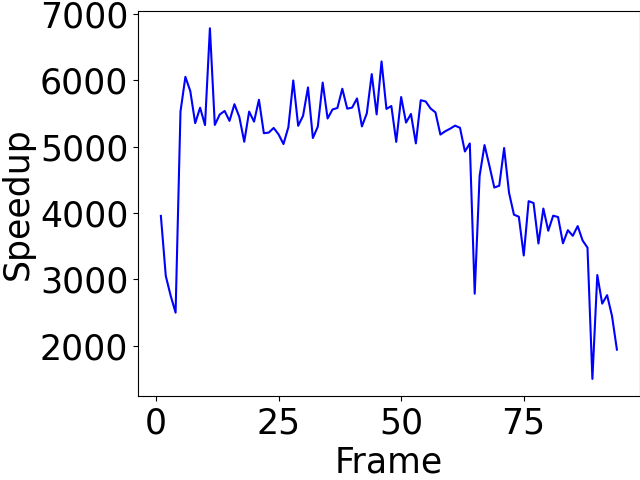}
            \caption{Cloth ball}
            \label{fig:5c}
        \end{subfigure}
        \hfill
        \begin{subfigure}[b]{0.24\textwidth}
            \includegraphics[width=\linewidth]{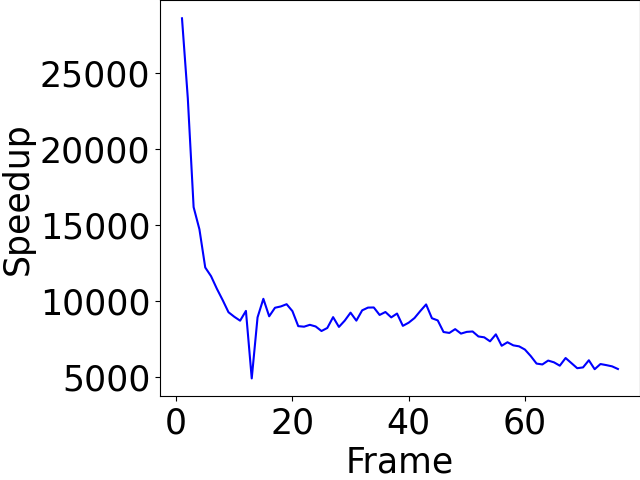}
            \caption{N-body}
            \label{fig:5d}
        \end{subfigure}
    \end{minipage}
    \caption{Per frame CD time comparison of Oibvh to Mochi}
    \label{fig:frames}
\end{figure*}

The table \ref{table:overall} also reports the time Oibvh spent in the broad phase in the same column as the average CD time.
We note that Mochi spends all of its CD time in the broad phase with only a negligible time in the narrow phase, so we do not report individual phase times.
Mochi achieves higher speedups on Flamenco and N-body datasets.
Observe that, for these particular datasets, Oibvh is speeding far lesser percentage of overall CD time in the broad phase.
Oibvh is spending more time in the narrow phase doing the time-consuming triangle-triangle intersection test.
Because of reduction \ref{red:4}, the result of Mochi's broad phase is either a triangle-triangle or ray-auxiliary-triangle intersection.
Hence, in the cases where Oibvh has to perform more triangle-triangle intersection tests, Mochi exhibits higher speedups.
Moreover, for the Flamenco dataset, where Mochi's speedup is the highest, Oibvh is detecting an unusually higher number of AABB intersections, which can be seen in table \ref{table:intersection}.
The CD time of Oibvh is rather proportional to the number of AABB intersections than the number of triangles in the scene.
For Mochi, we can not say the same as the number of broad phase intersections is also proportional to the number of triangles.

In figure \ref{fig:frames}, we further examine Mochi's CD performance by plotting per frame speedup of the CD time of Mochi over Oibvh.
For certain frames, we notice several orders of magnitude speedups,  especially for the middle frames.
The trend of speedup curves follows the motion of objects over the frames in the dataset.
The funnel dataset describes the motion of a ball wrapped in a cloth moving through a funnel with no collisions between the objects at the beginning and end. 
Hence, we see that speedups increase and decrease as the collisions increase and decrease, respectively.
The Flamenco dataset describes the interaction between layers of clothing as a dancer is moving, with layers of cloth apart from each other initially.
In the beginning, there are fewer to no collisions, so Oibvh does not spend any time performing triangle intersection tests.
But as frames progress, the number of intersections increases, making the speedups increase.
The cloth ball dataset contains the scene where a ball falls onto a cloth and gets wrapped in the cloth.
The N-body dataset describes a number of objects colliding and slowing down as time progresses.
So, we see a higher speedup at the beginning.


\section{Conclusion}
We show how to leverage RT cores for perform CD by building various reductions and implementing them.
Even though the BVH build by RT cores is inaccessible, Mochi cleverly launches rays to the trace the objects in the scene.
We developed object-object, coplanar ray-triangle and coplanar triangle-triangle intersection tests on a hardware that is limited to ray-AABB and non-coplanar ray-triangle intersection test.
Moreover, we do not employ any algorithmic optimizations and completely rely on our reductions to achieve several orders of magnitude speedups.
Our experiments revealed several insights about the RT cores' functioning.
For example, the BVH built by RT cores is not algorithmically superior to our baseline.
Our future work is to extend Mochi to include optimizations such as tracing a shared edge with single ray in reduction \ref{red:4}, and to support other types of CD such as continuous collision detection.

Though we are able to exploit RT cores to achieve high performance in case of CD, RT cores are still far from providing general-purpose acceleration.
They are limited in their capability to only build BVHs and traverse them.
While trying to reduce other applications to ray tracing is challenging, the end result of faster libraries justifies it.
We should explore the range of applications that can mapped to RT cores functionality as a community to benefit from the computing technologies available to us.

\subsection*{Acknowledgement}
We thank Nvidia Optix developer forum \cite{optixd} for being an inspiration in developing reduction \ref{red:4}. We are grateful for their suggestions.
\bibliographystyle{eg-alpha-doi}
\bibliography{references}

\newcommand{\etalchar}[1]{$^{#1}$}
\begin{thebibliography}{\uppercase{WTMT18}}

\bibitem[AMD23]{amd}
\textsc{AMD}:
\newblock Amd ray tracing, 2023.
\newblock URL: \url{https://www.amd.com/en/technologies/rdna}.

\bibitem[BA15]{breuer2015modeling}
\textsc{Breuer M., Almohammed N.}:
\newblock Modeling and simulation of particle agglomeration in turbulent flows
  using a hard-sphere model with deterministic collision detection and enhanced
  structure models.
\newblock \emph{International Journal of Multiphase Flow 73} (2015), 171--206.

\bibitem[CDK20]{oibvh}
\textsc{Chitalu F.~M., Dubach C., Komura T.}:
\newblock Binary ostensibly-implicit trees for fast collision detection.
\newblock In \emph{Computer Graphics Forum} (2020), vol.~39, Wiley Online
  Library, pp.~509--521.

\bibitem[CG98]{chakravarthy1998obstacle}
\textsc{Chakravarthy A., Ghose D.}:
\newblock Obstacle avoidance in a dynamic environment: A collision cone
  approach.
\newblock \emph{IEEE Transactions on Systems, Man, and Cybernetics-Part A:
  Systems and Humans 28}, 5 (1998), 562--574.

\bibitem[CMT02]{cordier2002real}
\textsc{Cordier F., Magnenat-Thalmann N.}:
\newblock Real-time animation of dressed virtual humans.
\newblock In \emph{Computer Graphics Forum} (2002), vol.~21, Wiley Online
  Library, pp.~327--335.

\bibitem[Cou15]{bullet}
\textsc{Coumans E.}:
\newblock Bullet physics simulation.
\newblock In \emph{ACM SIGGRAPH 2015 Courses}. 2015, p.~1.

\bibitem[DR11]{dehnen2011n}
\textsc{Dehnen W., Read J.~I.}:
\newblock N-body simulations of gravitational dynamics.
\newblock \emph{The European Physical Journal Plus 126} (2011), 1--28.

\bibitem[EPVV21]{evangelou}
\textsc{Evangelou I., Papaioannou G., Vardis K., Vasilakis A.~A.}:
\newblock Fast radius search exploiting ray tracing frameworks.
\newblock \emph{Journal of Computer Graphics Techniques (JCGT) 10}, 1 (February
  2021), 25--48.
\newblock URL: \url{http://jcgt.org/published/0010/01/02/}.

\bibitem[For23]{optixd}
\textsc{Forum N. O.~D.}:
\newblock Nvidia optix developer forum, 2023.
\newblock URL:
  \url{https://forums.developer.nvidia.com/t/ray-in-the-triangle-plane/261746}.

\bibitem[Int23]{intel}
\textsc{Intel}:
\newblock Intel ray tracing, 2023.
\newblock URL:
  \url{https://www.intel.com/content/www/us/en/developer/articles/guide/real-time-ray-tracing-in-games.html}.

\bibitem[LMM10]{lauterbach2010gproximity}
\textsc{Lauterbach C., Mo Q., Manocha D.}:
\newblock gproximity: hierarchical gpu-based operations for collision and
  distance queries.
\newblock In \emph{Computer Graphics Forum} (2010), vol.~29, Wiley Online
  Library, pp.~419--428.

\bibitem[M{\"o}l97]{moller1997fast}
\textsc{M{\"o}ller T.}:
\newblock A fast triangle-triangle intersection test.
\newblock \emph{Journal of graphics tools 2}, 2 (1997), 25--30.

\bibitem[NK23]{dbscan}
\textsc{Nagarajan V., Kulkarni M.}:
\newblock {RT-DBSCAN:} accelerating {DBSCAN} using ray tracing hardware.
\newblock \emph{CoRR abs/2303.09655} (2023).
\newblock URL: \url{https://doi.org/10.48550/arXiv.2303.09655}, \href
  {http://arxiv.org/abs/2303.09655} {\path{arXiv:2303.09655}}, \href
  {https://doi.org/10.48550/arXiv.2303.09655}
  {\path{doi:10.48550/arXiv.2303.09655}}.

\bibitem[NMK23]{trueknn}
\textsc{Nagarajan V., Mandarapu D., Kulkarni M.}:
\newblock Rt-knns unbound: Using {RT} cores to accelerate unrestricted neighbor
  search.
\newblock In \emph{Proceedings of the 37th International Conference on
  Supercomputing, {ICS} 2023, Orlando, FL, USA, June 21-23, 2023} (2023),
  Gallivan K.~A., Gallopoulos E., Nikolopoulos D.~S., Beivide R., (Eds.),
  {ACM}, pp.~289--300.
\newblock URL: \url{https://doi.org/10.1145/3577193.3593738}, \href
  {https://doi.org/10.1145/3577193.3593738}
  {\path{doi:10.1145/3577193.3593738}}.

\bibitem[nvi18]{nvidia_white}
\textsc{nvidia}:
\newblock Nvidia rt white paper, 2018.
\newblock URL:
  \url{https://images.nvidia.com/aem-dam/en-zz/Solutions/design-visualization/technologies/turing-architecture/NVIDIA-Turing-Architecture-Whitepaper.pdf}.

\bibitem[owl]{owl}
Owl: A node graph "wrapper" library for optix 7.
\newblock URL: \url{https://github.com/owl-project/owl}.

\bibitem[TMT10]{TMT10-GMOD}
\textsc{Tang M., Manocha D., Tong R.}:
\newblock Mccd: Multi-core collision detection between deformable models using
  front-based decomposition.
\newblock \emph{Graphical Models 72}, 2 (2010), 7--23.
\newblock \href {https://doi.org/DOI: 10.1016/j.gmod.2010.01.001}
  {\path{doi:DOI: 10.1016/j.gmod.2010.01.001}}.

\bibitem[TTS06]{tropp2006fast}
\textsc{Tropp O., Tal A., Shimshoni I.}:
\newblock A fast triangle to triangle intersection test for collision
  detection.
\newblock \emph{Computer Animation and Virtual Worlds 17}, 5 (2006), 527--535.

\bibitem[Wel13]{weller2013new}
\textsc{Weller R.}:
\newblock \emph{New geometric data structures for collision detection and
  haptics}.
\newblock Springer Science \& Business Media, 2013.

\bibitem[WTMT18]{wang2018}
\textsc{Wang X., Tang M., Manocha D., Tong R.}:
\newblock Efficient bvh-based collision detection scheme with ordering and
  restructuring.
\newblock In \emph{Computer graphics forum} (2018), vol.~37, Wiley Online
  Library, pp.~227--237.

\bibitem[WUM{\etalchar{*}}19]{wald}
\textsc{Wald I., Usher W., Morrical N., Lediaev L., Pascucci V.}:
\newblock {RTX Beyond Ray Tracing: Exploring the Use of Hardware Ray Tracing
  Cores for Tet-Mesh Point Location}.
\newblock In \emph{High-Performance Graphics - Short Papers} (2019),
  Steinberger M., Foley T., (Eds.), The Eurographics Association.
\newblock \href {https://doi.org/10.2312/hpg.20191189}
  {\path{doi:10.2312/hpg.20191189}}.

\bibitem[YL07]{yershova2007improving}
\textsc{Yershova A., LaValle S.~M.}:
\newblock Improving motion-planning algorithms by efficient nearest-neighbor
  searching.
\newblock \emph{IEEE Transactions on Robotics 23}, 1 (2007), 151--157.

\bibitem[Zhu22]{rtnn}
\textsc{Zhu Y.}:
\newblock Rtnn: Accelerating neighbor search using hardware ray tracing.
\newblock In \emph{Proceedings of the 27th ACM SIGPLAN Symposium on Principles
  and Practice of Parallel Programming} (New York, NY, USA, 2022), PPoPP '22,
  Association for Computing Machinery, p.~76–89.
\newblock URL: \url{https://doi.org/10.1145/3503221.3508409}, \href
  {https://doi.org/10.1145/3503221.3508409}
  {\path{doi:10.1145/3503221.3508409}}.

\bibitem[ZWW20]{zellman}
\textsc{Zellmann S., Weier M., Wald I.}:
\newblock Accelerating force-directed graph drawing with rt cores.
\newblock In \emph{2020 IEEE Visualization Conference (VIS)} (2020),
  pp.~96--100.
\newblock \href {https://doi.org/10.1109/VIS47514.2020.00026}
  {\path{doi:10.1109/VIS47514.2020.00026}}.

\end{thebibliography}

\end{document}